\documentclass[conference]{IEEEtran}
\IEEEoverridecommandlockouts
\usepackage{cite}
\usepackage{amsmath,amssymb,amsfonts}
\usepackage{algorithmic}
\usepackage{graphicx}
\usepackage{textcomp}
\usepackage{xcolor}
\usepackage{bm}
\usepackage{optidef}
\usepackage{scalerel}
\usepackage[labelfont=bf]{caption}
\def\BibTeX{{\rm B\kern-.05em{\sc i\kern-.025em b}\kern-.08em
    T\kern-.1667em\lower.7ex\hbox{E}\kern-.125emX}}

\begin{document}

\title{DSROQ: Dynamic Scheduling and Routing for QoE Management in LEO Satellite Networks\\
}

% \author[1]{Dhiraj Bhattacharjee}
% \author[1]{Pablo G. Madoery}
% \author[2]{Abhishek Naik}
% \author[3]{Halim Yanikomeroglu}
% \author[4]{\\G{\"{u}}ne{\c{s}} Karabulut Kurt}
% \author[5]{St\'ephane Martel}
% \author[5]{Khaled Ahmed}

% \affil[1,3]{\normalsize Department of Systems and Computer Engineering, Carleton University, Canada}
% \affil[2]{\normalsize National Research Council Canada, Canada}
% \affil[4]{\normalsize Department of Electrical Engineering, Polytechnique Montréal, Canada}
% \affil[5]{\normalsize Satellite Systems, MDA, Canada}

% % Emails aligned with author order, avoiding OCR issues
% \affil[ ]{\normalsize dhirajbhattacharjee@sce.carleton.ca, pablomadoery@sce.carleton.ca, abhishek.naik@nrc-cnrc.gc.ca,}
% \affil[ ]{\normalsize halim@sce.carleton.ca, gunes.kurt@polymtl.ca, stephane.martel@mda.space, khaled.ahmed@mda.space}

\author{
\IEEEauthorblockN{
Dhiraj Bhattacharjee\IEEEauthorrefmark{1},
Pablo G. Madoery\IEEEauthorrefmark{1},
Abhishek Naik\IEEEauthorrefmark{2},
Halim Yanikomeroglu\IEEEauthorrefmark{1},\\
Gunes Karabulut Kurt\IEEEauthorrefmark{3},
St\'ephane Martel\IEEEauthorrefmark{4},
Khaled Ahmed\IEEEauthorrefmark{4}
}
\IEEEauthorblockA{\IEEEauthorrefmark{1} Department of Systems and Computer Engineering, Carleton University, Canada}
\IEEEauthorblockA{\IEEEauthorrefmark{2}National Research Council Canada, Canada}
\IEEEauthorblockA{\IEEEauthorrefmark{3} Department of Electrical Engineering, Polytechnique Montréal, Canada}
\IEEEauthorblockA{\IEEEauthorrefmark{4} Satellite Systems, MDA, Canada}
}

\maketitle

\begin{abstract}
The modern Internet supports diverse applications with heterogeneous quality of service (QoS) requirements. Low Earth orbit (LEO) satellite constellations offer a promising solution to meet these needs, enhancing coverage in rural areas and complementing terrestrial networks in urban regions. Ensuring QoS in such networks requires joint optimization of routing, bandwidth allocation, and dynamic queue scheduling, as traffic handling is critical for maintaining service performance. This paper formulates a joint routing and bandwidth allocation problem where QoS requirements are treated as soft constraints, aiming to maximize user experience. An adaptive scheduling approach is introduced to prioritize flow-specific QoS needs. We propose a Monte Carlo tree search (MCTS)-inspired method to solve the NP-hard route and bandwidth allocation problem, with Lyapunov optimization-based scheduling applied during reward evaluation. Using the Starlink Phase 1 Version 2 constellation, we compare end-user experience and fairness between our proposed DSROQ algorithm and a benchmark scheme. Results show that DSROQ improves both performance metrics and demonstrates the advantage of joint routing and bandwidth decisions. Furthermore, we observe that the dominant performance factor shifts from scheduling to routing and bandwidth allocation as traffic sensitivity changes from latency-driven to bandwidth-driven.
\end{abstract}

\vspace{-0.5em}

\section{Introduction}
The global push for universal Internet access has intensified with the rise of 6G, which promises ultra-fast, reliable, and versatile communication. A key challenge is bridging the digital divide—the gap between those with and without access to high-speed internet. Satellites are vital to this effort, especially in remote regions lacking terrestrial infrastructure, and in urban areas with high demand. Non-terrestrial networks (NTNs), led by industry players like SpaceX, Telesat, Amazon, and OneWeb, are gaining momentum in both industry and academia, in particular, low Earth orbit (LEO) satellite constellations due to their lower latency and higher data rates.

Today’s Internet supports diverse applications from VoIP and video conferencing (VC) to live streaming (LS), messaging, and file transfers (FT), each with distinct quality of service (QoS) needs defined by latency, throughput, packet loss, and jitter. While QoS refers to measurable network performance metrics, quality of experience (QoE) is the user-perceived performance. Moreover, flows from different applications may carry varying priorities based on service level agreements (SLAs). Meeting these heterogeneous QoS and SLA requirements to optimize QoE depends on two critical decisions: (i) routing and bandwidth allocation at the network level, and (ii) queue management at the node level. Optimizing either in isolation may fall short. Thus, a joint approach that combines both is crucial for delivering optimal user experience.

Early queue management techniques like first-in-first-out (FIFO), priority queuing (PQ), and weighted fair queuing (WFQ) were designed to provide simplicity, prioritization for critical traffic, and fair bandwidth allocation, respectivey. Building on these, integrated services (IntServ) \cite{intserv} introduced per-flow resource reservations for strict QoS but lacked scalability, prompting the development of differentiated services (DiffServ) \cite{diffserv}, which classifies traffic as expedited forwarding (EF), assured forwarding (AF), and best effort (BE). Active queue management (AQM) techniques emerged to mitigate congestion proactively. Random early detection (RED) \cite{red} drops packets probabilistically to avoid buffer overflows but required careful tuning whereas controlled delay (CoDel) \cite{codel} improves latency by dropping packets exceeding a delay threshold with minimal configuration. Flow queue CoDel (FQ-CoDel) \cite{fqcodel} improves fairness and delay control by combining CoDel with per-flow queuing, making it effective for diverse and modern traffic environments.  

Effective queue management enhances traffic handling within the network, but optimal routing strategies are equally crucial for meeting diverse QoS requirements and maximizing overall performance. In \cite{10233775}, a multi-level priority queuing scheme is proposed where priorities adapt dynamically based on delay estimation and QoS violations. \cite{QoSRA} builds on this by combining priority queuing with weighted round-robin (WRR) scheduling for three traffic classes. \cite{10390619} uses bandwidth and delay metrics for route selection with FIFO queues. \cite{efficientdiffrouting} sequentially computes routes for delay-sensitive, bandwidth-sensitive, and BE flows using Dijkstra’s algorithm for minimum latency, maximum bandwidth, and fewest hops, respectively. Finally, \cite{fybrrlink} proposes fybrrLink, which uses flow-specific weights (latency, bandwidth, reliability) to compute a modified-cost shortest path via a constrained search space generated using a modified Bresenham’s algorithm.

While extensive research exists on QoS-aware routing in satellite networks, most works do not assume specific queuing architectures despite queuing delay being a key factor influencing routing decisions. Some studies incorporate priority queuing or combine it with WRR schedulers, but the absence of joint routing and node-level scheduling decisions often leads to suboptimal resource utilization. In parallel, the combinatorial complexity of joint routing across large-scale flows has led many recent approaches to adopt sequential methods, which compromise overall network performance. Moreover, modern queue management techniques primarily emphasize per-hop behavior, ignoring routing context and end-to-end network conditions. Traditional models also treat QoS constraints rigidly, deeming any violation infeasible. A more flexible strategy considers QoS as soft objectives, aiming to minimize violations or maximize satisfaction. To the best of our knowledge, no prior work jointly optimizes routing, bandwidth allocation, and queue scheduling while accounting for diverse traffic priorities and QoS requirements. We address this gap by proposing a unified optimization framework across all flows that maximizes the weighted sum of end-user experience.

The paper organization is as follows. Section~\ref{sec:network_arq} introduces the network architecture. Section~\ref{sec:problem_form} formulates the joint routing and bandwidth allocation problem along with dynamic queue scheduling problem to maximize overall user experience. Section~\ref{sec:dsroq} presents a tractable solution using Monte Carlo tree search (MCTS) for routing and bandwidth allocation, combined with a Lyapunov optimization-based queue scheduling algorithm. Section~\ref{sec:result} compares the proposed approach with a benchmark scheme. Finally, Section~\ref{sec:conclusion} concludes the paper and outlines directions for future work.\vspace{-0.5em}

\section{Network Architecture}\label{sec:network_arq}

We consider a LEO satellite network with regenerative payloads, where satellites are interconnected with up to four bidirectional inter-satellite links (ISLs) to their immediate neighbors. The network is represented as a graph $(\mathcal{V}, \mathcal{E})$, where $\mathcal{V}$ and $\mathcal{E}$ denote the sets of satellites and ISLs, respectively. All ISLs have the same unidirectional link capacity, $R_l$. Ground traffic generation and reception are abstracted by modeling both source and destination of a flow $\Tilde{f}$ as satellites, i.e., the satellite serving a ground area acts as the source or destination of flows from/to that area. Each flow is identified by a unique ID, source and destination satellite, and application type. To improve scalability and reduce per-flow control complexity, $\beta$ flows with the same source, destination, and application type are aggregated into a single aggregated flow $f$. To capture the dynamics of mobile LEO satellites, we adopt the virtual node model from \cite{virtualnode}, where each satellite is associated with a fixed ground area it serves. As the topology evolves, the next satellite in orbit assumes responsibility for that area. To enable seamless transitions, routing information is transferred to the incoming satellite prior to handover, ensuring continuity in routing decisions.

Each aggregated flow has QoS requirements defined by upper and lower bounds on latency ($\delta_f^{\max}$, $\delta_f^{\min}$), throughput ($\tau_f^{\max}$, $\tau_f^{\min}$), and packet drop rate ($l_f^{\max}$, $l_f^{\min}$), which influence the end-user experience. Let $\mathcal{F}$ denote the set of $F$ aggregated flows, partitioned into EF, AF, and BE classes: $\mathcal{F} = \mathcal{F}_{EF} \cup \mathcal{F}_{AF} \cup \mathcal{F}_{BE}$. Each aggregated flow is assigned an SLA-based weight $w_f$. We define the time slot $t$ as the time needed to transmit a packet, i.e., packet length divided by link capacity. Packets arriving via ISLs or generated at a satellite are processed by a routing module. If the satellite is not the destination, the routing module determines the next hop, and the packet is forwarded to one of four queuing modules. Each queuing module classifies packets by their aggregated flow ID and stores them in dedicated per-aggregated-flow queues. A scheduler then selects one aggregated flow to transmit its head-of-queue (HoQ) packet over the link. The scheduling index $s_{f,\{i,j\}}^{t}$ is a binary variable set to 1 if the aggregated flow $f$ is scheduled at time slot $t$ on the outbound ISL $\{i,j\}$ from satellite $i$ to $j$, and 0 otherwise. For packet dropping due to buffer overflow, packets are considered to be dropped in round robin fashion from the aggregated flows which share a common outbound link.
\section{Problem Formulation}\label{sec:problem_form}
To evaluate the performance of each aggregated flow $f$ at its destination, we measure average latency ($\delta_f^T$), throughput ($\tau_f^T$), and packet drop rate ($l_f^T$) over a time window $T$ comprising $Z>1$ time slots. Based on these metrics, we compute a QoS score reflecting user experience. The best and worst QoS scores are denoted by $\Omega^{\max}$ and $\Omega^{\min}$, respectively. Latency-based QoS score $\Omega_f^{\delta}$ is assigned as $\Omega^{\max}$ if $\delta_f^T \leq \delta_f^{\min}$, and $\Omega^{\min}$ if $\delta_f^T \geq \delta_f^{\max}$. For intermediate values, $\Omega_f^{\delta}$ is linearly interpolated between $\Omega^{\max}$ and $\Omega^{\min}$. A similar mapping is applied to the average packet drop rate $l_f^T$ to compute the QoS score $\Omega_f^{l}$, using the bounds $l_f^{\min}$ and $l_f^{\max}$. For throughput, the QoS score $\Omega_f^{\tau}$ is computed inversely to latency: $\Omega^{\max}$ is assigned when $\tau_f^T \geq \tau_f^{\max}$, and $\Omega^{\min}$ when $\tau_f^T \leq \tau_f^{\min}$. For intermediate values, $\Omega_f^{\tau}$ is linearly interpolated between $\Omega^{\min}$ and  $\Omega^{\max}$. The final QoS score $\Omega_f^T$ is computed as a weighted sum of the individual scores:
$\Omega_f^T = \zeta_f^\delta \Omega_f^\delta + \zeta_f^\tau \Omega_f^\tau + \zeta_f^l \Omega_f^l$,
where $\zeta_f^\delta$, $\zeta_f^\tau$, and $\zeta_f^l$ are the respective weights for latency, throughput, and packet drop rate for the aggregated flow $f$.\vspace{-0.5em}
\subsection{Route and Bandwidth Allocation}
Let $K_f$ denote the number of available routes for aggregated flow $f$ from source to destination. We define a binary vector $\bm{\alpha}_f^T$, where each element $\alpha_{f,r}^T \in \{0,1\}$ indicates whether route $r$ is selected for aggregated flow $f$ during time window $T$. The selected route for aggregated flow $f$ with $\alpha_{f,r}^T = 1$ is denoted by $r_f^T$, representing an ordered list of edges from source to destination. Upon route selection, the allocated bandwidth $B_f^T$ must be determined as well. The bandwidth $B_f^T$ is chosen from a discrete set $\mathcal{B}_f$ of $B_w$ values uniformly spanning $[\tau_f^{\min}, \tau_f^{\max}]$, including both endpoints. Our objective is to maximize the weighted sum of the QoS scores of all the aggregated flows $f\in\mathcal{F}$ scaled to 1 as presented below:\vspace{-1.5em}

\begin{maxi!}|s|
    {\scaleto{\alpha_{f,r}^T,B_f^T}{10pt}}{\frac{\sum_{f\in\mathcal{F}}\;w_f\Omega_f^T}{\Omega^{\max}\sum_{f\in\mathcal{F}}\;w_f}}{}{}\label{p1}    
  \addConstraint{\sum_{r=1}^{K_f}\alpha_{f,r}^T\;=\;1,\;\forall f\in\mathcal{F}}\label{p1c1}
    \addConstraint{\alpha_{f,r}^T\in \{0,1\},\;\forall f\in\mathcal{F},r=1,2,...,K_f}\label{p1c2}
    \addConstraint{B_{f}^T\in \mathcal{B}_f,\;\forall f\in\mathcal{F}}{}\label{p1c3}
    \addConstraint{\sum_{f:\{i,j\}\in r_f^T} B_f^T\;\leq\;R_l,\;\forall \{i,j\}\in\mathcal{E.}}\label{p1c6}
\end{maxi!}

\vspace{-0.3em}The maximization in (\ref{p1}) selects optimal routes $\alpha_{f,r}^T$ and bandwidths $B_f^T$ for all aggregated flows. Constraint (\ref{p1c1}) ensures each flow selects one route among $K_f$ options, while (\ref{p1c2}) enforces the binary nature of $\alpha_{f,r}^T$. (\ref{p1c3}) restricts $B_f^T$ to discrete values in $\mathcal{B}_f$, and (\ref{p1c6}) ensures that the total bandwidth on any link does not exceed its capacity $R_l$. The problem is an integer non-linear program (INLP) due to the non-linear objective, and is NP-hard \cite{hartmanis1982computers}. To address this with tractable complexity, we propose a scalable MCTS-based joint routing and bandwidth allocation scheme, detailed in the next section.
\subsection{Queue Scheduling}
With the given routes and bandwidth allocations for all aggregated flows, the scheduler must determine, at each time slot, which aggregated flow to serve from the set of aggregated flows traversing the link $\{i,j\}$. Since throughput can be evaluated on a per-hop basis, we estimate the latency-based QoS score at each hop and each time slot, denoted by $\Omega_{f,\{i,j\}}^{\delta,t}$, as follows:  
\vspace{-0.5em}
\begin{equation}\label{new_qos}
    \Omega_{f,\{i,j\}}^{\delta,t} = \Omega^{\delta} \big( \delta_{f,\{i,j\}}^t, \delta_{f,\{i,j\}}^{\max}, \delta_{f,\{i,j\}}^{\min} \big),
\end{equation}
where $\Omega^{\delta}(\cdot)$ denotes the delay-to-QoS mapping function. Here, $\delta_{f,\{i,j\}}^t$ represents the local queuing delay of the HoQ packet at satellite $i$. The parameters $\delta_{f,\{i,j\}}^{\max}$ and $\delta_{f,\{i,j\}}^{\min}$ are modified delay bounds used to estimate the latency-based QoS score. These modified bounds are obtained by subtracting, from the original end-to-end delay bounds, the sum of the propagation delay, $\delta_{f,\{i,j\}}^{prop}$ and transmission delay, $\delta_{f,\{i,j\}}^{trans}$ of the selected route $r_f^T$ scaled by the number of hops $|r_f^T|$. The resulting per-hop queuing delay bound is further adjusted by the congestion factor $\gamma_{\{i,j\}} / \overline{\gamma_f}$, where $\gamma_{\{i,j\}}$ denotes the average link utilization of ISL $\{i,j\}$ and $\overline{\gamma_f}$ is the average link utilization over the entire route of aggregated flow $f$. For example, the modified upper latency bound is given by:  
\vspace{-0.56em}
\begin{equation}\label{new_threshold}
    \delta_{f,\{i,j\}}^{\max} =
    \frac{\delta_f^{\max} - \sum_{\{i,j\} \in r_f^T} \big( \delta_{f,\{i,j\}}^{prop} + \delta_{f,\{i,j\}}^{trans} \big)}
    {|r_f^T|} \cdot \frac{\gamma_{\{i,j\}}}{\overline{\gamma_f}}.
\end{equation}

In (\ref{new_threshold}), the inclusion of the congestion factor adjusts the bound relative to the route-wide average link utilization of the aggregated flow’s route, stretching it for links whose average link utilization exceeds this route-wide average and compressing it for links whose utilization is lower, thereby accounting for localized congestion in scheduling decisions. The weighted estimated latency-based QoS score can be written as,
\begin{equation}\label{objective}
    g_{f,\{i,j\}}^t=w_f\zeta_f^{\delta}\Omega^{\delta}\Big(\delta_{f,\{i,j\}}^t+1-s_{f,\{i,j\}}^{t}, \delta_{f,\{i,j\}}^{max}, \delta_{f,\{i,j\}}^{min}\Big)
\end{equation}
(\ref{objective}) represents that the local HoQ delay is in terms of time slots and gets incremented by one if not scheduled. To represent the throughput deficit of an aggregated flow, we define a concatenated vector of virtual queues, $\bm{V}_{\{i,j\}}^t=[V_{f,\{i,j\}}^t];\;\forall f:\{i,j\}\in r_f^T$ and each virtual queue is updated as follows:\vspace{-0.4em}
\begin{equation}
    V_{f,\{i,j\}}^{t+1}=max[V_{f,\{i,j\}}^t+B_f^T-s_{f,\{i,j\}}^{t},0].
\end{equation}

With this, we formulate the following scheduling problem for each outbound ISL $\{i,j\}$ considering $N$ time slots:\vspace{-1.5em}

\begin{maxi!}|s|
    {\scaleto{s_{f,\{i,j\}}^{t}}{10pt}}{\lim_{N\to \infty}\frac{1}{N}\sum_{t=1}^N\sum_{f:\{i,j\}\in r_f^T}g_{f,\{i,j\}}^t}{}{}\label{p2}    
  \addConstraint{\lim_{N\to \infty}\frac{\mathbb{E}\big[V_{f,\{i,j\}}^t\big]}{N}=0,\forall f:\{i,j\}\in r_f^T,\forall t}\label{p2c1}
    \addConstraint{s_{f,\{i,j\}}^{t}\in \{0,1\},\forall f:\{i,j\}\in r_f^T,\forall t}\label{p2c2}
    \addConstraint{\sum_{f:\{i,j\}\in r_f^T}s_{f,\{i,j\}}^{t}=1,\forall t.}\label{p2c3}
\end{maxi!}
The problem (\ref{p2}) determines the optimal scheduling index, $s_{f,\{i,j\}}^{t}$, $\forall t$ for the aggregated flows with routes passing through ISL $\{i,j\}$. Constraint (\ref{p2c1}) ensures mean rate stability of $V_{f,\{i,j\}}^t$, i.e., ensuring the throughput to be stabilized around the assigned bandwidth, $B_f^T$. Finally, (\ref{p2c2}) and (\ref{p2c3}) ensures the binary nature of the scheduling index and scheduling only one aggregated flow per time slot, respectively.
\section{Dynamic Scheduling and ROuting for QoE Management (DSROQ)}\label{sec:dsroq}
\subsection{MCTS-based Routing and Bandwidth Allocation}
MCTS, introduced by Coulom \cite{coulom2006efficient}, is a popular algorithm for solving combinatorial and sequential decision problems, especially in large and uncertain state spaces. MCTS builds a search tree through four iterative steps: selection (choosing nodes by balancing exploration and exploitation), expansion (adding new child nodes), simulation (estimating outcomes via random rollouts), and backpropagation (updating node values based on simulation results). This iterative process enables efficient discovery of high-reward solutions. A detailed description of these steps follows.
\subsubsection{Tree Structure}
Each training episode starts from an empty network at the root node $S_0=\{\}$. At each tree level, beginning with $S_0$, a configuration of route and bandwidth, $\{\bm{\alpha}_f^T, B_f^T\}$ is selected for an aggregated flow $f$ and appended to the current node to form a child node. For example, after selecting $\{\bm{\alpha}_1^T, B_1^T\}$ at level 1 for aggregated flow 1, the transition $S_0 \rightarrow S_{11}$ yields $S_{11} = \{\bm{\alpha}_1^T, B_1^T\}$. At level 2, choosing $\{\bm{\alpha}_2^T, B_2^T\}$ leads to $S_{11} \rightarrow S_{21}$, where $S_{21} = \{\{\bm{\alpha}_1^T, B_1^T\}, \{\bm{\alpha}_2^T, B_2^T\}\}$. Continuing this process across flows and levels constructs the complete route and bandwidth allocations upon reaching a leaf node.

\subsubsection{Selection, Expansion, and Simulation}
Starting from $S_0$, if the current node has no children, a random rollout is performed, selecting random configurations until reaching a leaf node. Otherwise, it applies an $\epsilon$-greedy strategy to either select an existing child node or expand a new one.

\subsubsection{Reward}

At the end of each episode, upon reaching a leaf node with the full configuration $\prod_{f\in\mathcal{F}}\bm{\alpha}_f^T, B_f^T$, the objective function (\ref{p1}) is computed based on the QoS scores. To account for constraint violation, a cost function $C$ is defined for (\ref{p1c6}), while constraints (\ref{p1c1})–(\ref{p1c3}) are inherently satisfied through discrete decision variables, as shown below:\vspace{-0.5em}

\begin{equation}\label{c2} C=\sum_{\{i,j\}\in\mathcal{E}}\frac{\max(0,\sum_{f:\{i,j\}\in r_f^T} B_f^T-R_l)}{R_l}.
\end{equation}

Here, $C$ penalizes bandwidth allocations exceeding link capacity $R_l$. The final reward can be written as,\vspace{-0.5em}

\begin{equation}\label{rwd} 
\mathcal{R}= \frac{\sum_{f\in\mathcal{F}}\;w_f\Omega_f^T}{\Omega^{\max}\sum_{f\in\mathcal{F}}\;w_f}-\lambda C,
\end{equation}
where $\lambda$ balances the trade-off between objective maximization and constraint satisfaction.
\subsubsection{Backpropagation}Each node maintains a Q-value representing its selection quality. After completing an episode, the computed reward $\mathcal{R}$ from (\ref{rwd}) is backpropagated from the leaf node to the root through the connecting edges. At each node $S$, the Q-value is updated as,\vspace{-0.9em}

\begin{equation}\label{updateQ} Q(S) \xleftarrow{}\max\{Q(S),\mathcal{R}\}. \end{equation}

Finally, the $\epsilon$ value is updated at each episode using the following logarithmic function of the number of episodes, starting at $(1 - \epsilon_0)$ and gradually decreasing to $\epsilon^{\min}$:\vspace{-1.4em}

\begin{equation}\label{epsillor_modify}
    \epsilon\Big|_{episode=z}=\max\{1-\epsilon_0*\log_{10}(\frac{z}{a_0}+b_0),\epsilon^{\min}\}.
\end{equation}

Over multiple episodes, the MCTS policy learns to identify the near-optimal path from the root node to one of the leaf nodes yielding the highest reward encountered during training.\vspace{-0.5em}

\subsection{Lyapunov Optimization based Dynamic Scheduling}
In (\ref{p2}), to handle the stochastic constraint (\ref{p2c1}), using Lyapunov optimization, the Lyapunov function of the concatenated vectors are defined as,
\begin{equation}
    L[\bm{V}_{\{i,j\}}^t]\triangleq \frac{1}{2} \sum_{f:\{i,j\}\in r_f^T}\Big(V_{f,\{i,j\}}^t\Big)^2, \forall t,
\end{equation}
while the conditional Lyapunov drift is defined as,
\begin{equation}
    \Delta(\bm{V}_{\{i,j\}}^t)\triangleq \mathbb{E}\Big[L[\bm{V}_{\{i,j\}}^{t+1}]-L[\bm{V}_{\{i,j\}}^t]\Big|\bm{V}_{\{i,j\}}^t\Big].
\end{equation}
From \cite{neely2010stochastic}, the upper bound of $\Delta(\bm{V}_{\{i,j\}}^t)$ can be written as,
\begin{equation}
    \Delta(\bm{V}_{\{i,j\}}^t)\leq B+\sum_{f:\{i,j\}\in r_f^T}\mathbb{E}\Big[V_{f,\{i,j\}}^t(B_f^T-s_{f,\{i,j\}}^{t})\Big|\bm{V}_{\{i,j\}}^t\Big],
\end{equation}
where $B$ is a scalar constant. The overall objective is maximizing (\ref{p2}) while minimizing the upper bound of $\Delta(\bm{V}_{\{i,j\}}^t)$, which leads to the following objective function: 
\begin{equation}\label{driftpluspenalty}
    G^t_{\{i,j\}}=\sum_{f:\{i,j\}\in r_f^T}g_{f,\{i,j\}}^t-w_f\zeta_f^{\tau}\Delta(\bm{V}_{\{i,j\}}^t).
\end{equation}
Now, putting the upper bound of the conditional Lyapunov drift in (\ref{driftpluspenalty}), and only keeping the relevant terms, the modified objective is presented as follows:
\begin{equation}\label{finalobjective}
    \begin{split}        
    G^{t^\prime}_{\{i,j\}}= & \!\!\!\!\sum_{f:\{i,j\}\in r_f^T}\!\!\!\!w_f\Big\{\zeta_f^{\delta}\Omega^{\delta}\Big(\delta_{f,\{i,j\}}^t+1-s_{f,\{i,j\}}^{t}\Big) \\
    & +\zeta_f^{\tau}V_{f,\{i,j\}}^t s_{f,\{i,j\}}\Big\}. 
    \end{split}
\end{equation}
After mathematical manipulation, (\ref{finalobjective}) can be solved with constraints (\ref{p2c2}) and (\ref{p2c3}), and the optimal aggregated flow $f^*$ at the egress port of ISL $\{i,j\}$ at time slot $t$ with $s_{f,\{i,j\}}^{t}=1$ can be written as follows:
\begin{equation}\label{schedulingrule}
\begin{split}
        f^*=\arg \max_{f:\{i,j\}\in r_f^T} & w_f\Big\{\zeta_f^{\delta}\Big[\Omega^{\delta}\Big(\delta_{f,\{i,j\}}^t\Big)-\Omega^{\delta}\Big(\delta_{f,\{i,j\}}^t+1\Big)\Big]\\
    &+\zeta_f^{\tau}V_{f,\{i,j\}}^t\Big\}.
\end{split}
\end{equation}\vspace{-1em}

DSROQ integrates the MCTS-based policy for route and bandwidth allocation with the scheduling policy defined in (\ref{schedulingrule}). During each MCTS training episode, when a leaf node is reached, the reward is evaluated using (\ref{rwd}), based on the average network performance over a time window of $Z$ time slots. Within this window, and for every time slot, the packet to be transmitted is selected using the scheduling policy (\ref{schedulingrule}). Thus, the scheduling policy directly influences the reward estimation in MCTS, creating a tightly coupled loop between long-term path decisions and short-term queue management.\vspace{-0.5em}
\section{Simulation Results}\label{sec:result}
\subsection{Simulation Settings}
In this section, we present and compare the training and testing results with fybrrLink \cite{fybrrlink}, which calculates routes sequentially for each incoming flow. We evaluate fybrrLink under a prioritized scheme where EF flows are served first, followed by AF and BE. Two algorithm variants are tested: DSROQ and DSROQ-FIFO. Both use MCTS-learned route and bandwidth allocations; DSROQ applies the scheduling rule in (\ref{schedulingrule}), while DSROQ-FIFO uses a FIFO queue per egress port toward each outbound ISL. We calculate propagation delay for intra and inter-orbital ISLs in a 4-by-4 satellite configuration, using the Starlink Phase 1 Version 2 constellation ($550$ km altitude, $53^\circ$ inclination). The simulation runs for $60$ seconds with topology changes every $15$ seconds. We simulate $10,000$ flows across VC (EF), LS (AF), video-on-demand (AF), and FT (BE) applications with distributions of $20\%$, $20\%$, $20\%$, and $40\%$, respectively for $100$ iterations. Packet generation for each flow is modeled as a Poisson random variable with a mean equal to its allocated bandwidth. EF, AF and BE flow weights are considered as $20$, $2$ and $1$, respectively unless stated otherwise. The latency, bandwidth, and packet loss related QoS bounds are considered from \cite{cisco}, \cite{fcc}, and \cite{qospdr}, respectively. In addition, $[\zeta_f^{\delta}, \zeta_f^{\tau},\zeta_f^{l}]$ for EF, AF, and BE flows are considered as [0.8,0.1,0.1], [0.1,0.45,0.45], and [0.1,0.8,0.1], respectively \cite{qosweights}. Finally, $\Omega_{max}$ and $\Omega_{min}$ are considered as 5 and 1, respectively. \vspace{-0.5em}
\subsection{Reward and Constraint Violation}
\begin{figure}
    \centering
    \includegraphics[width=0.7\linewidth]{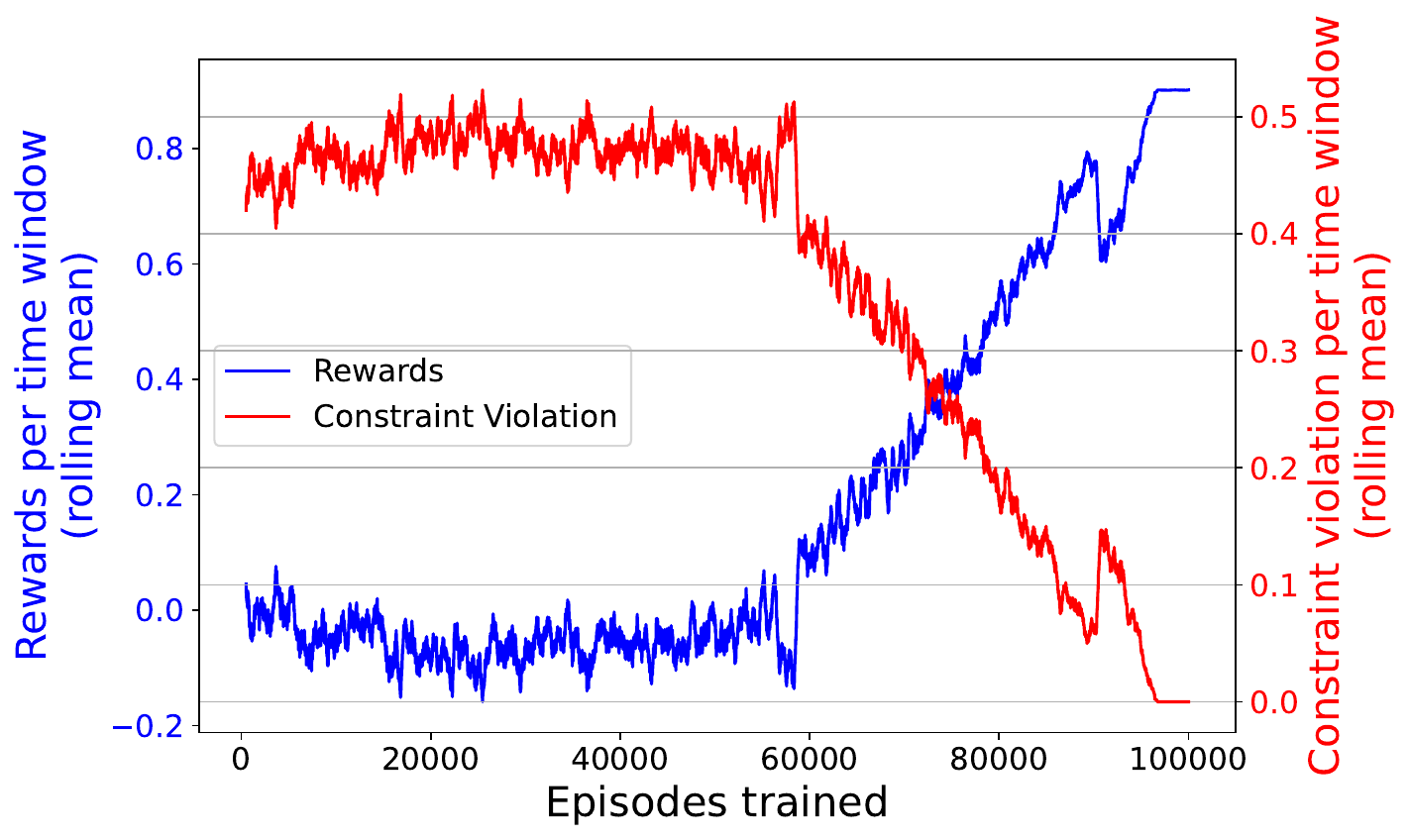}
    \caption{Reward and constraint violation vs episodes trained.\vspace{-1em}}
    \label{fig:rwd}
\end{figure}
Fig. \ref{fig:rwd} shows the rolling mean (window size 500) of reward and constraint violations per time window versus training episodes. Reward values increase and eventually saturate as training progresses, driven by the decay of the exploration factor $\epsilon$ toward $\epsilon^{\min}=0$ as per (\ref{epsillor_modify}). Similarly, constraint violations decrease and converge to zero, indicating successful training and constraint compliance.
\subsection{User Experience}
\begin{figure}
    \centering
    \includegraphics[width=0.6\linewidth]{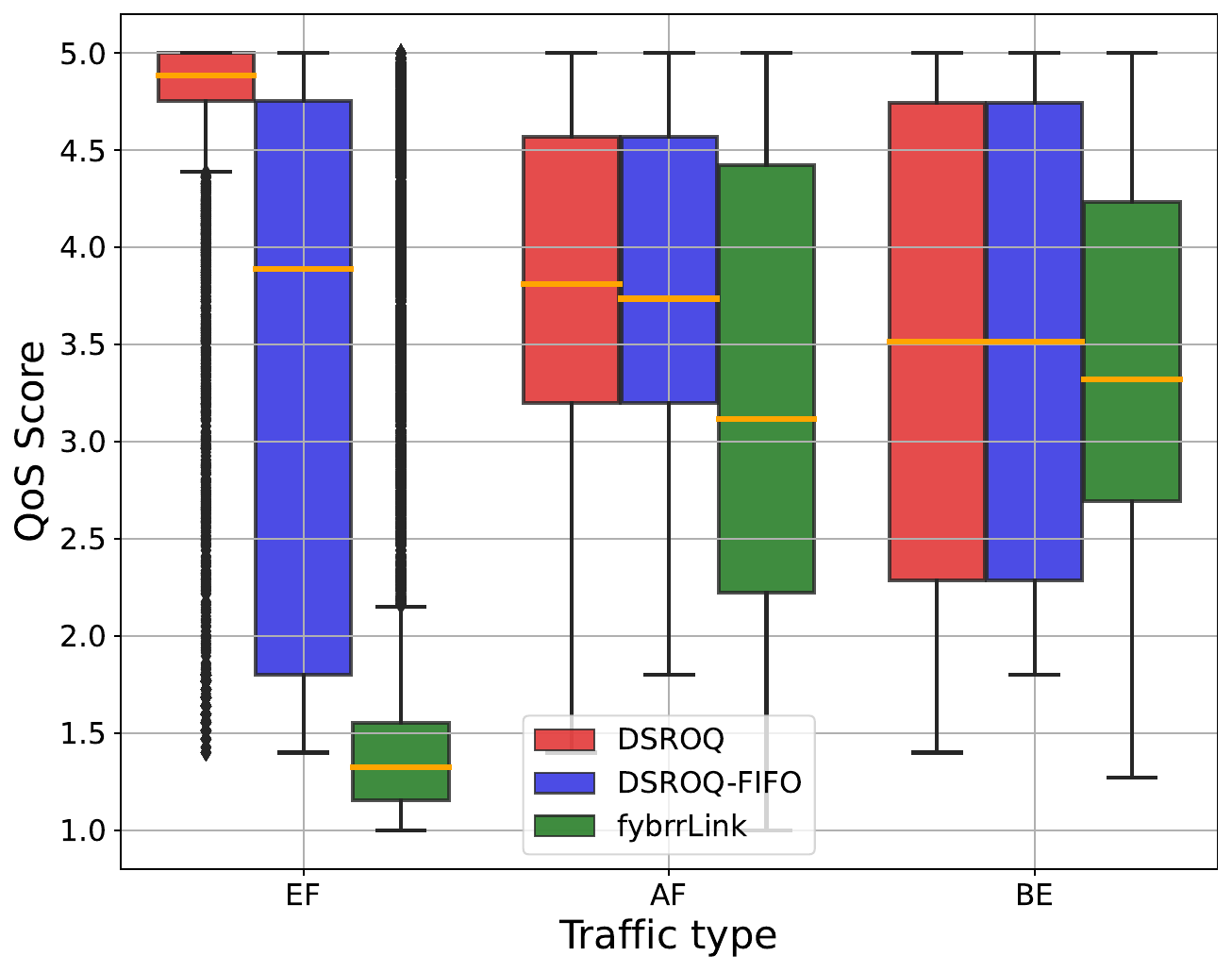}
    \caption{Comparison of QoS scores.\vspace{-1em}}
    \label{fig:qoe}
\end{figure}
Fig. \ref{fig:qoe} compares the QoS scores of the proposed DSROQ algorithm against the benchmark for EF, AF, and BE traffic classes. The box plots show that the median QoS score (yellow lines) decreases across DSROQ variants from EF to AF to BE flows, reflecting the priority weights used during training and testing (EF $>$ AF $>$ BE). In contrast, fybrrLink exhibits an increasing median QoS trend from EF to BE flows, as QoS requirements become progressively relaxed. Compared to the benchmark, DSROQ achieves significant QoS improvements for EF and AF flows, with the improvement diminishing toward BE flows. Although several outliers are visible in the DSROQ EF scores, they represent around 4\% of the total data and do not significantly affect the overall distribution. Comparing DSROQ and DSROQ-FIFO reveals a large difference for EF flows in terms of median QoS scores and inter-quartile ranges, which reduces for AF and becomes negligible for BE. First of all, changing the scheduling policy from (\ref{schedulingrule}) to a FIFO rule removes the priority in scheduling EF packets and gives equal chance to an AF or BE packet to get scheduled. This leads to a latency increase in EF flows but latency reduction in AF and BE flows. Secondly, when the average incoming traffic is below link capacity, scheduling mainly impacts delay rather than throughput. Since EF flows are highly delay-sensitive ($\zeta_f^{\delta}/\zeta_f^{\tau}=8$), change in scheduling policy has a strong effect in overall QoS distribution for EF flows, whereas for AF ($\zeta_f^{\delta}/\zeta_f^{\tau}=0.22$) and BE ($\zeta_f^{\delta}/\zeta_f^{\tau}=0.125$) flows, the impact is progressively smaller.\vspace{-0.5em}
\subsection{Effect of Priority Weights}
\begin{figure}
    \centering
    \includegraphics[width=0.6\linewidth]{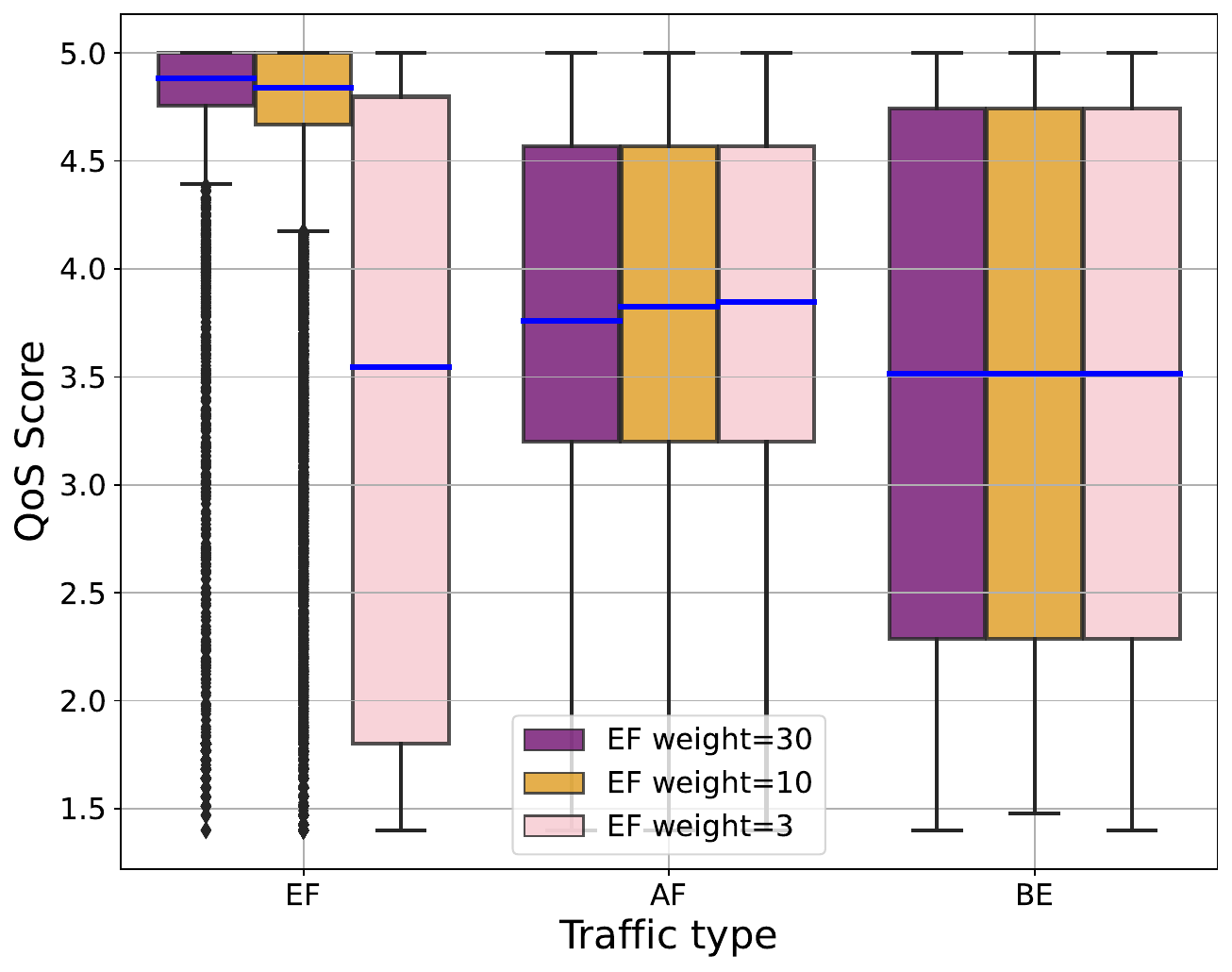}
    \caption{Effect on varying priority weights on QoS scores.}
    \label{fig:efweight}
\end{figure}
Fig. \ref{fig:efweight} shows the QoS score variation across three EF priority weights, with route and bandwidth allocations trained for an EF weight of 20 (AF and BE weights of 2 and 1, respectively used in both training and testing). For high EF weights, the median QoS scores (blue lines) follow EF$>$AF$>$BE. However, when the EF weight is lesser, this order changes. Reducing the EF weight lowers the scheduling priority of EF packets, increasing their latency and degrading QoS for latency-sensitive EF flows. This reduces median QoS scores for EF flows along with increase in outlier percentage from 1.6\% to 14.3\% for EF weight of 20 and 10. Meanwhile, AF and BE packets are scheduled more often, slightly improving AF QoS (due to moderate latency sensitivity) and leaving BE QoS largely unchanged (due to minimal latency sensitivity). These results highlight that throughput is mainly determined by route and bandwidth allocations, while latency is influenced by scheduling, route, and bandwidth allocations.
\begin{figure}
    \centering
    \includegraphics[width=0.6\linewidth]{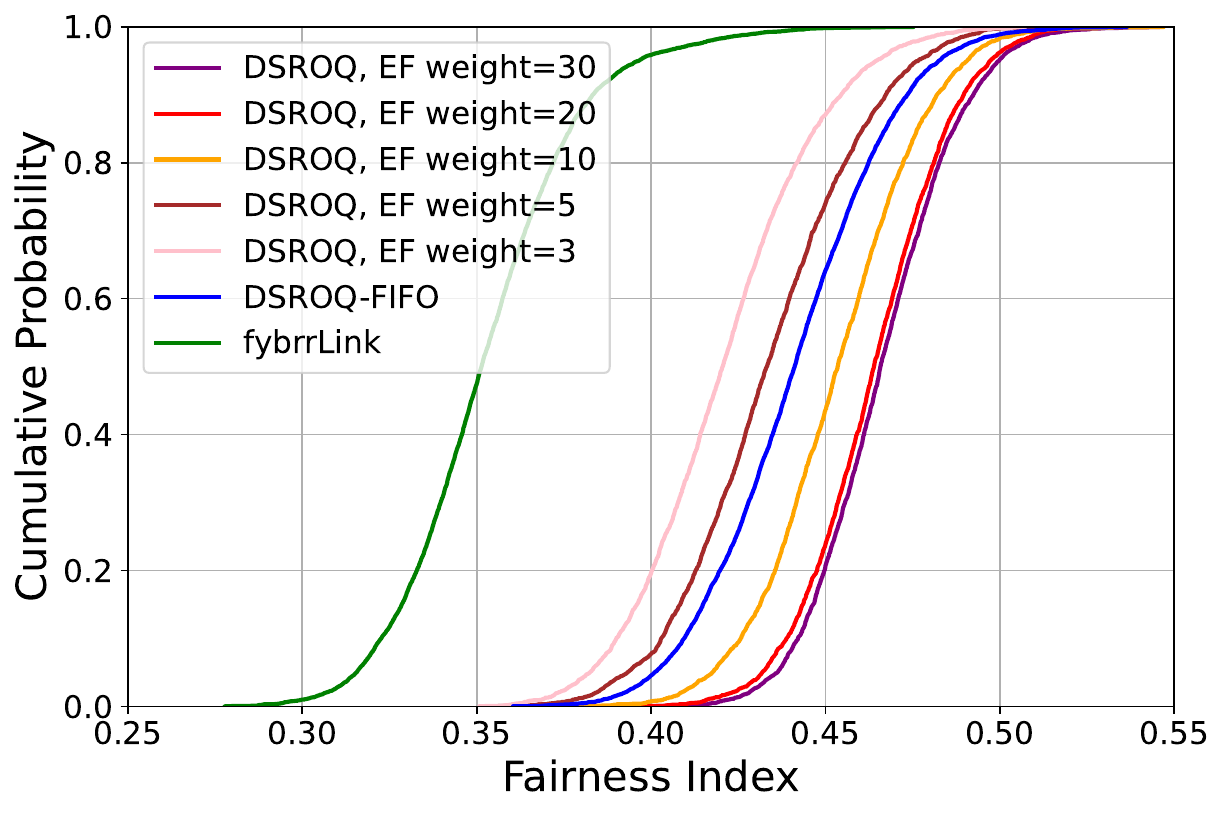}
    \caption{CDF of fairness.}
    \label{fig:fairness}
\end{figure}
\subsection{Fairness Index}
Fig. \ref{fig:fairness} presents a comparative analysis of the cumulative distribution function (CDF) of fairness index between the benchmark algorithm and proposed DSROQ under different EF priority weights using the route and bandwidth allocations trained with an EF weight of 20 (AF and BE weights fixed at 2 and 1, respectively). Fairness of the QoS scores provided to each aggregated flow is measured using the QoE fairness index from \cite{QoEfairness}. We observe that DSROQ consistently achieves higher fairness than fybrrLink, primarily due to its joint route and bandwidth allocation that considers all flows simultaneously, as opposed to fybrrLink’s sequential processing. In addition, DSROQ with EF weight of 20 also outperforms DSROQ-FIFO in fairness, which aligns with Fig. \ref{fig:qoe}, where the inter-quartile range of EF QoS scores increases from DSROQ to DSROQ-FIFO, indicating higher QoS variance and lower fairness. Furthermore, increasing the EF weight improves fairness among flows in DSROQ. This is supported by Fig. \ref{fig:efweight}, which shows that higher EF weights reduce the spread in EF QoS scores, thereby increasing fairness.
\section{Conclusion and Future Work}\label{sec:conclusion}
Meeting heterogeneous QoS requirements and service-level priorities in satellite networks relies on routing, bandwidth allocation, and scheduling decisions. This paper models end-user experience as QoS scores derived from network performance metrics, and formulates the network performance as a weighted sum of QoS scores, where weights reflect service-level priorities. To maximize this metric, we pose an optimization problem for jointly selecting routes and bandwidth allocations, along with an adaptive scheduling problem that accounts for flow-specific QoS needs and priorities. An MCTS-inspired approach is proposed to address the NP-hard problem of route and bandwidth allocation, with Lyapunov optimization-based scheduling solution used during reward evaluation in MCTS. The resulting algorithm, DSROQ, combines both the MCTS-based resource allocation and the proposed scheduler. We evaluate two DSROQ variants against the benchmark fybrrLink in a 4-by-4 satellite grid within the Starlink Phase 1 Version 2 constellation. Both DSROQ variants outperform fybrrLink in QoS score and fairness. DSROQ-FIFO’s improvement over fybrrLink highlights the benefit of joint routing and bandwidth allocation across all flows, while DSROQ’s advantage over DSROQ-FIFO emphasizes the critical role of scheduling, especially for latency-sensitive EF flows. This effect is further validated by varying EF priority weights, where the dominant influence shifts from scheduling to routing and bandwidth allocation as flow sensitivity moves from latency to bandwidth.

While this paper optimizes the network over a fixed time window for multiple flows, future extensions may consider dynamic networks where traffic arrives and departs randomly. In such settings, resource allocation could either react to available capacity on arrival or perform joint optimization periodically to accommodate all active flows. In future, we plan to extend this framework to support joint admission control, routing, bandwidth allocation, and scheduling for such dynamic and diverse traffic scenarios.\vspace{-0.5em}

\bibliographystyle{IEEEtran}
\footnotesize
% \vspace{-0.5em}
\bibliography{name}

\end{document}